\documentstyle[12pt,epsfig]{article}
\def \ut#1{\rlap{\lower1ex\hbox{$\sim$}}#1{}}
\begin{document}
\title{Regularization of the Hamiltonian constraint and the closure of the
constraint algebra} \author{Roumen Borissov\thanks{e-mail address:
borissov@astro.ocis.temple.edu}
\\Physics Department , Temple University\\ Philadelphia, PA 19122}
\maketitle

\begin{abstract}

In the paper we discuss the process of regularization of the
Hamiltonian constraint in the Ashtekar approach to quantizing gravity.
We show in detail the calculation of the action of the regulated Hamiltonian
constraint on Wilson loops. An important issue considered in the paper is the
closure of the constraint algebra. The main result we obtain is that the
Poisson bracket
between the regulated Hamiltonian constraint and the Diffeomorphism constraint
is equal to a sum of regulated Hamiltonian constraints with appropriately
redefined regulating functions.

\end{abstract}
\newpage

\section{Introduction}

The recent progress, made toward quantizing gravity has been mostly due to
the introduction of a new set of canonically conjugate variables by Ashtekar
\cite{Ash1}. In terms of the Ashtekar variables the gravitational constraints
in the canonical approach have acquired simple polynomial form. It has allowed
for different solutions to the quantum gravitational constraints to be
investigated in the connection representation \cite{Smo1}, and in the loop
representation \cite{Rov1}. Successful steps have been
performed also toward incorporating scalar \cite{SmoRov} and fermion
\cite{Rov3} matter fields in the theory. Regardless of the success of the
Ashtekar approach there are still a lot of difficulties in giving precise
physical meaning to  the obtained results. An important problem in the theory
is the regularization of the products of distributional quantities  involved
in the calculations. The need for regularization in the Ashtekar approach was
encountered by Jacobson and Smolin  in their paper on the connection
representation \cite{Smo1}. Since then it has attracted a lot of attention.
Thorough discussions of the regularization procedure is given in \cite{Tsamis}
and \cite{Jack}. The regularization has been applied in \cite{Pull},
\cite{Blen}, \cite{Gam2}  when the action of the Hamiltonian constraint on the
wave functions of the theory has been investigated.  In \cite {Smo3} detailed
discussion has been given to the result from regularization for the action of
the local operator representing the metric. In \cite{SmoRov} the regularization
has played major role for the successful inclusion of the matter fields in the
theory. Recently in \cite{Gam1} the authors have spelled in details the
calculations of the constraints algebra purely in the loop representation.

In the present paper we are concerned mostly with a precise definition of the
regularization procedure and with consequences of regularization for the
coordinate invariance and the constraints closure. To avoid unnecessary
complications we work in the more conventional connection representation. The
content of the paper is as follows: In  section 2. we justify the necessity
for regularization considering the action of the unregularized gravitational
constraints on the wave functions of the theory. After that in section 3. we
introduce the regularization procedure and calculate in detail the action of
the regulated Hamiltonian constraint.
Despite the fact that such type of calculations have been performed by
Blencowe \cite{Blen} in the loop representation, we have been encouraged by
the results obtained in \cite{SmoRov} to investigate again the apparent
background dependance. As we will see an undesired imprint does indeed survive
in the contribution from smooth loops and loops with kinks but, confirming the
result from \cite{SmoRov}, we show that the background dependance drops
completely from the contribution from self-intersections. In section 4. we
investigate the constraints algebra at classical and quantum level, paying
close attention to the problem for the algebra closure. On contrary to some
previous results \cite{Blen} we show that the Poisson bracket between the
regulated Hamiltonian and the Diffeomorphism constraints equals a sum of
regulated Hamiltonian constraints. We conclude with discussion of some open
questions. Some technical details of the computations are included in
appendices.

\section{Ashtekar variables, Wilson loops, and necessity for regularization}

The classical phase space of canonical quantum gravity { \it a la}
Ashtekar consists of a configuration variable
$A_{a}^{i}$, which is a complex $SU(2)$ connection on the spatial manifold,
and its conjugate momentum
$\tilde{E}^{a}_{i}$, a triad with density weight one. (As usual:
$a$,$b$,... are spatial indices; $i$,$j$,... are internal indices; each tilde
denotes density weight one.)
In terms of Ashtekar variables the constraints of the theory
have the form:

\[
\tilde{{\cal G}}_{i}  =  {\cal D}_{a}\widetilde{E}_{i}^{a}\cong 0
\]
\[
\tilde{{\cal V}}_{a}  = F_{ab}^{i}\widetilde{E}^{b}_{i}\cong 0
\]
\[
 \stackrel{\approx}{{\cal H}}  =  {1\over 2}\epsilon_{ijk}F_{ab}^{k}
\widetilde{E}^{ai}
\widetilde{E}^{bj}\cong 0,
\]
where ${\cal D}_{a}$ and $F_{ab}^{i}$ are correspondingly  the covariant
derivative and the curvature defined with the Ashtekar connection $A_{a}^{i}$.
The constraints are the so called Gauge constraint, Vector constraint
and Hamiltonian constraint. The first constraint generates rotations
in the internal space, reflecting the freedom of choosing
different $``\widetilde{E}"$-s representing the same
3-metric. The second constraint modulo the first one generates
spatial diffeomorphism transformations. The Hamiltonian
constraint governs the evolution of the system under consideration
from one space slice into another.
As usual we will smear out the
constraints with some arbitrary appropriately densitized fields:

\begin{equation}\label{Gs}
{\cal G}({\bf \vec{N}})= \int d^3x {\bf N}^{i}(x){\cal D}_{a}
\widetilde{E}_{i}^{a}\cong 0
\end{equation}

\begin{equation}\label{Vs}
{\cal V}(\vec N) =
\int d^3x N^{a}(x)F_{ab}^{i}(x)\widetilde{E}^{b}_{i}(x) \cong 0
\end{equation}

\begin{equation}\label{Hs}
{\cal H}(\ut{N})= \int d^3x\ut{N}(x)
\epsilon_{ijk}F_{ab}^{k}(x)\widetilde{E}^{ai}(x)
\widetilde{E}^{bj}(x)\cong 0.
\end{equation}

We also will write in explicit form the constraint
${\cal C}(\vec{N})$ which generates spatial diffeomorphism transformations:

\[
{\cal C}(\vec{N}) = {\cal V}(\vec N) - {\cal G}({N^{a} \vec{A}_{a}}) =
\int d^3x N^{a}(x)[F_{ab}^{i}(x)\widetilde{E}^{b}_{i}(x) - A^{i}_{a}
 {\cal D}_{b}{\tilde E}^{b}_{i}]\cong 0.
\]
The Poisson bracket of this constraint with any function $f(A,E)$ of the
canonical variables gives:

\begin{equation}\label{Lf}
\{ {\cal C}(\vec{N}) , f(A,E)\} = {\cal L}_{\vec N} f(A,E),
\end{equation}
where ${\cal L}_{\vec N}$ is the Lie derivative along the field ${\vec N}$.

The classical constraint algebra is closed, that is the Poisson brackets
between constraints give as a result combinations of the constraints
themselves. The exact expressions are:

\[
\{{\cal G}({\bf \vec{M}}), {\cal G}({\bf \vec{N}})\} = i{\cal G}
({\bf \vec{M}}\times {\bf \vec{N}}),
\]

\[
\{{\cal C}({ \vec{N}}),{\cal G}({\bf \vec{M}})\} = {\cal G}
( {\cal L}_{\vec{N}}{\bf \vec{M}}),
\]

\[
\{{\cal G}({\bf \vec{M}}), {\cal H}(\ut{N})\} = 0,
\]

\[
\{{\cal C}({ \vec{N}}), {\cal C}({ \vec{M}})\} = {\cal C}
({\cal L}_{\vec{N}}{ \vec{M}}),
\]

\begin{equation}\label{C,H}
\{{\cal H}(\ut{N}) , {\cal C}({ \vec{M}})\} = - {\cal H}
({\cal L}_{\vec{N}}\ut{N}),
\end{equation}
and
\begin{equation}\label{H,H}
\{{\cal H}(\ut{N}), {\cal H}(\ut{M})\} =
{\cal C}(\vec{K}) + {\cal G}(K^{a}\vec{A}_{a}) =
{\cal V}(\vec{K})
\end{equation}
with
\begin{equation}\label{K}
K^{a}(x) = \tilde{E}^{aj}(x)\tilde{E}^{c}_{j}(x)
[\ut{M}(x)\partial_{c}\ut{N}(x) -
\ut{N}(x)\partial_{c}\ut{M}(x)].
\end{equation}

In the process of quantization, the canonical variables become operators of
multiplication and differentiation and  the Poisson brackets become
commutators. We also have to choose a particular factor ordering for
expressions containing operator products (see \cite{Pull} for discussion of
the problem). In the constraints we will put all the ``$\tilde{E}$"-s to the
right of the ``$A$"-s. The reason we choose such a factor ordering is the fact
that  with this ordering a set of solutions to all quantum constraints has
been found. The solution obtained is given by regular functionals of knot and
link classes of smooth, non-intersecting loops. However there is a problem
with the chosen factor ordering: The Poisson bracket of two Hamiltonian
constraints is nontrivial - on the right hand side we still have a combination
of the constraints but the coefficients are not constants. They are functions
of the basic variables. With the chosen factor ordering  both ``$E$"-s from
(7) will appear to the right of the Vector constraint in (\ref{H,H}) and this
prevents the quantum constraint algebra from closing.

In the proposed paper we will work in the more conventional connection
representation in which the wave functionals $ \Psi(A)$ depend on the Ashtekar
connection. The loop and the connection representations are connected via
the (not yet fully justified, see {\cite {Ash3}}) loop transform:

\[
\Psi(\gamma)= \int{``d\mu[A]"}W_{ \gamma}(A)\Psi(A).
\]
In this Fourier-like transformation $``d\mu[A]"$ is an unknown measure on the
space of connections modulo gauge transformations ${\cal A}/{\cal G}$. The
kernel of the transformation $W_{ \gamma}(A)$ is an Wilson loop. The Wilson
loops form an infinite basis of gauge invariant functionals, parametrized by a
loop $\gamma$. They are defined by the trace of the path ordered
exponential of the line integral of the connection along the loop $\gamma$:

\begin{equation}\label{PsiA}
W_{ \gamma}(A) = { \rm Tr} U\left( 0,1 \right) = { \rm Tr} \left( {\cal P}
{ \rm exp} \oint{ ds { \dot \gamma}^{a}(s) A_{a}^{i} ( \gamma(s)) } \tau_{i}
\right),
\end{equation}
where

\[
U\left( s_{1},s_{2} \right)={ \rm Tr} \left( {\cal P} { \rm exp}
\int^{ s_{2}}_{ s_{1}}{ ds { \dot \gamma}^{a}(s) A_{a} ( \gamma(s)) } \right)
\]
is the holonomy of $A$ along the loop $\gamma$ and $\tau_{i}=-{i \over
2}\sigma_{i}$; $\sigma_{i}$ are the Pauli matrices. Because of the loop
transform
the investigation of the action of constraints on the wave functionals in the
loop representation is equivalent to considering the corresponding action in
the connection representation on the Wilson loops.

By their construction as a trace of holonomy, the Wilson loops are
automatically gauge invariant.
The action of the Diffeomorphism constraint on the Wilson loop is given by:

\[
{\hat {\cal C}}(\vec N )W_{ \gamma}(A) =
\int{d^{3}x N^{a}(x)F_{ab}^{i}(x)
\oint{ds \delta^{3}({\tilde x},\gamma(s))\dot{\gamma}^{b}(s){ \rm Tr}[U(0,s)
\tau_{i}U(s,1)]}}=
\]

\begin{equation}\label{C-ill}
= \oint{ds}\dot{\gamma}^{b}(s)N^{a}(\gamma(s))F_{ab}^{i}
(\gamma(s)){ \rm Tr}[U(0,s)\tau_{i}U(s,1)].
\end{equation}
Similarly we will get for the action of the Hamiltonian constraint:

\[
{\hat {\cal H}}(\ut{M})W_{ \gamma}(A) =
\]

\begin{equation}\label{H-ill}
= \int{d^{3}x}\oint{ds} \oint{dt}
\dot{\gamma}^{a}(s)\dot{\gamma}^{b}(t) \ut{M}(x)\epsilon_{ijk}F_{ab}^{k}(x)
\delta^{3}({\tilde x},\gamma(s)) \delta^{3}({\tilde x},\gamma(t))
L^{ij}(s,t;\gamma),
\end{equation}
where with $L^{ij}(s,t;\gamma)$ we have denoted the ``loop deformation" - the
result of the action of the Hamiltonian constraint on the loop itself. This
action amounts to breaking the loop and inserting Pauli matrices at certain
points. The analytic expression for $L^{ij}(s,t;\gamma) $ is:

\[
L^{ij}(s,t;\gamma) = {\rm Tr}[ \vartheta(s-t)U(0,t)\tau^{j}U(t,s)\tau^{i}
U(s,1) +
\]

\begin{equation}\label{Lij}
+ \vartheta(t-s)U(0,s)\tau^{i}U(s,t)\tau^{j}U(t,1)].
\end{equation}
Expression (\ref {H-ill}) above is ill defined and requires regularization.
The problem arises from the fact that it contains two spatial
$\delta $-functions integrated in a 5-fold integral. In the next section we
will consider the point-splitting regularization as a possible way for solving
the problem. This procedure will make the action of ${\hat {\cal H}}(\ut{M})$
well
defined and after appropriate renormalization - finite. Also we will
investigate the consequences of the regularization procedure for the closure
of the constraint algebra.

\section{Regulated Hamiltonian constraint and its action on Wilson loops}
\label{action}
\subsection{How the regulator should look like?}

In the process of regularization we use
the following expression for $ {\cal H}(\ut{M})$:
\begin{equation}\label {regH}
{\cal H}_{\epsilon}(\ut{M}) = \int
d^{3}x\epsilon_{ijk}\ut{M}(x)F_{ab}^{k}(x)\int
d^{3}y  f_{\epsilon}(\tilde{x},y)
\tilde{E}^{ai}(y)\int
d^{3}z  f_{\epsilon}(\tilde{x},z)\tilde{E}^{bj}(z).
\end{equation}

In this expression $f_{\epsilon}(\tilde{x},y)$ is a regulating function
depending on a continuous parameter $\epsilon$. Here the tilde over $x$ means
that $f_{\epsilon}(\tilde{x},y)$ is a density with respect to its first
argument.
In \cite{SmoRov} Smolin and Rovelli have used regularization with
such symmetric
point-splitting in the loop representation and their result of the action of
the regulated Hamiltonian constraint is background independent.

The regulating function satisfies the requirement that for any
smooth function $\phi (x)$:

\begin{equation}
\lim_{\epsilon\rightarrow 0}{\int{d^{3}x \phi (x)f_{\epsilon}(\tilde{x},y)}}
= \phi (y).
\end{equation}

Particular examples of such functions are a normalized, weighted $\vartheta$-
function:

\begin{equation}\label{theta}
f_{\epsilon}(\tilde{x},y)=\sqrt{h(x)}f_{\epsilon}(x,y) =(3/4\pi\epsilon^{3})
\sqrt{h(x)}\vartheta
[ \epsilon -|{\vec x} - { \vec y}|]
\end{equation}
or Gaussian function

\[
f_{\epsilon}(\tilde{x},y)=(\epsilon\sqrt{\pi})^{-3}\sqrt{h(x)}\exp[ -
{ {|{\vec x} - { \vec y}|^{2}}\over {\epsilon}^{2}}].
\]

In the above expressions $h(x)$ is the determinant of the (arbitrary, i.g.
Euclidian) background metric $h^{ab}(x)$. This metric is used also in the
definition of the distance $|\vec{x} - \vec{y}|$. The process of regularization
amounts to performing all of the calculations in the action of
$\hat{{\cal H}}_{\epsilon}(\ut{M})$ on a Wilson loop and then taking the
limit  $\epsilon\rightarrow0$.
As we will see the problem of multiplying distributional quantities reduces to
the emergence of a single pole in $\epsilon$. Thus by renormalizing this last
expression we will get a well defined, finite result.

Using (\ref {regH}) we get for the action of the regulated Hamiltonian
constraint on the Wilson loop:
\[
\hat{\cal H}_{\epsilon}(\ut{M})W_{ \gamma}(A) =  \int{d^{3}x}\oint{ds}
\oint{dt}
\dot{\gamma}^{a}(s)\dot{\gamma}^{b}(t) \ut{M}(x) \epsilon_{ijk}F_{ab}^{k}(x)
\]

\[
\int{d^{3}y}\int{d^{3}z}\delta^{3}(\tilde y,\gamma(s)) \delta^{3}(\tilde z,
\gamma(t))
f_{\epsilon}(\tilde{x},y)f_{\epsilon}(\tilde{x},z)L^{ij}(s,t;\gamma)
\]

\begin{equation}\label{Hpsi}
= \int{d^{3}x}\oint{ds} \oint{dt}
\dot{\gamma}^{a}(s)\dot{\gamma}^{b}(t) \ut{M}(x)\epsilon_{ijk}F_{ab}^{k}(x)
f_{\epsilon}(\tilde{x},\gamma(s))f_{\epsilon}(\tilde{x},\gamma(t))L^{ij}(s,t;
\gamma),
\end{equation}
where $L^{ij}(s,t;\gamma)$ is defined with (\ref {Lij}).
We will use the  $\vartheta$-function as a regulator in our calculations.
Combining (\ref {theta}) with (\ref {Hpsi}) we get:

\[
\hat{\cal H}_{\epsilon}(\ut{M})W_{ \gamma}(A) =  {9 \over
16\pi^{2}\epsilon^{6}} \oint{ds} \oint{dt}
\dot{\gamma}^{a}(s)\dot{\gamma}^{b}(t)L^{ij}(s,t;\gamma)
\]

\begin{equation}\label{Hpsi2}
\int{d^{3}x}\ut{M}(x) \epsilon_{ijk}F_{ab}^{k}(x)h(x)
\vartheta [ \epsilon - |\vec{x} - \vec{\gamma(s)}|]\vartheta [ \epsilon -
|\vec{x} - \vec{\gamma(t)}|].
\end{equation}

Because of the $\vartheta$-functions the last integral is non-vanishing only
for values of $ x $ in the region which is the intersection of the spheres
with radii $ \epsilon $ and centers $ \gamma(s) $ and  $ \gamma(t) $
correspondingly. The calculations from now on will depend on the type of the
loop which parametrizes the functional $W_{ \gamma}(A)$. In this paper we will
consider single smooth loops, loops with self-intersections and loops with
kinks.

\subsection{Smooth portions of loops}
\begin{figure}
\par
\centerline{
\epsfig{figure=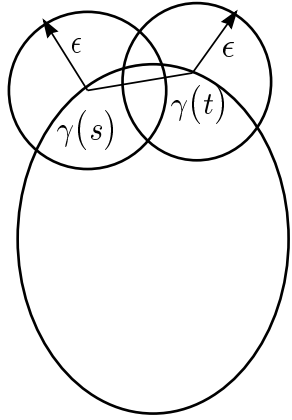,width=40mm}
}
\par
\caption{Smooth loop.}
\label{f1}
\end{figure}

We will start with the calculation of the contribution from smooth portions
of the loop $\gamma$. On Figure \ref{f1} we have shown single smooth loop but
our
calculations will also be valid for cases when the loop $\gamma$ has kinks or
self-intersection.

{}From Figure \ref {f1} we can determine that the distance between the centers
of the spheres
$\delta(s,t)$ is
at most twice the (small) parameter $\epsilon$, which enables us to make some
approximations.
Let us first consider the expansion:

\[
\delta(s,t)=|\sum_{n=1}^{\infty}{\vec\gamma}^{(n)}(s){(t-s)^{n} \over n!}|
\leq |\dot{\vec \gamma}||t-s| + |\sum_{n=2}^{\infty}{\vec\gamma}^{(n)}(s)
{(t-s)^{n} \over n!}| \leq 2\epsilon.
\]
{}From here we can determine the range of one of the parameters along the loop,
say $t$ with respect to $s$:

\[
t\in[  s -{2\epsilon\delta^{-} \over |\dot{\vec \gamma}|} , s + {2\epsilon
\delta^{+} \over |\dot{\vec \gamma}|}],
\]
where $\delta^{-}=1+{\cal O}(\epsilon)$ and $\delta^{+}=1+{\cal O}(\epsilon)$.
This means that we can fix $s$ in (\ref{Hpsi2}) and expand all functions of $t$
about $s$ in powers of $\epsilon$. The first term in the expansion of
$\dot{\gamma}^{b}(t)$ will give us the product:

\[
\dot{\gamma}^{a}(s)\dot{\gamma}^{b}(s)F_{ab}^{k},
\]
which, because of the antisymmetry of $ F_{ab}^{k}$, will make the whole
integral vanishing. Thus the first non-vanishing term will come from the
expansion of:

\[
{9 \over
16\pi^{2}\epsilon^{6}} \oint{ds} \oint{dt}
\dot{\gamma}^{a}(s)\ddot{\gamma}^{b}(s)(t-s)L^{ij}(s,t;\gamma)
\]

\begin{equation}\label{Hpsi3}
\int{d^{3}x}\ut{M}(x) \epsilon_{ijk}F_{ab}^{k}(x)h(x)
\vartheta [ \epsilon - |\vec{x} - \vec{\gamma(s)}|]\vartheta [ \epsilon -
|\vec{x} - \vec{\gamma(t)}|].
\end{equation}
In the expansion of the holonomies in $L^{ij}(s,t;\gamma)$ we have to keep
only terms of  zeroth order. Also, because the last integral in  (\ref{Hpsi3})
is different from zero only in a region of linear size $\epsilon$ we can
replace this integral with its mean value:

\[
\ut{M}(x_{0})F_{ab}^{k}(x_{0})\sqrt{h(x_{0})} \int{d^{3}x}\sqrt{h(x)}
\vartheta [ \epsilon - |\vec{x} - \vec{\gamma(s)}|]\vartheta [ \epsilon -
|\vec{x} - \vec{\gamma(t)}|]=
\]

\[
\ut{M}(x_{0})F_{ab}^{k}(x_{0})\sqrt{h(x_{0})} V(\delta(s,t)),
\]
where $V(\delta(s,t))$ is the volume of the intersection region,
$\delta(s,t)=|\vec{\gamma(s)} - \vec{\gamma(t)}|$ and $x_{0}$ is a point in
close vicinity to $\vec{\gamma(s)}$ and $\vec{\gamma(t)}$. The volume of the
intersection of two spheres with radii $\epsilon$ and distance between their
centers $\delta$ is equal to twice the volume cut from a sphere by a plane
passing at a distance $\delta/2$ from the center of the sphere. The volume of
the intersection is given by the expression:

\begin{equation}\label{Volume}
V(\delta)={ 2 \over 3 }\pi\epsilon^{3}  \left[ 2 - { 3 \over 2 }{ \delta
\over \epsilon} + \left( {\delta \over 2\epsilon} \right)^{3} \right].
\end{equation}

Because of the reparametrization invariance of the integrals in $s$ and $t$ we
can use a parametrization in which $|\dot{\vec \gamma}|=1$. After expanding
and performing the integration with respect to $t$ we get to the lowest order
in $\epsilon$ (see Appendix 1):

\[
\hat{\cal H}_{\epsilon}(\ut{M})W_{ \gamma}(A) =
\]

\[
{3\over 2\pi\epsilon}((\delta^{+})^{2}+(\delta^{-})^{2}) \oint{ds}
\dot{\gamma}^{a}(s)\ddot{\gamma}^{b}(s)M(\gamma(s)) F_{ab}^{k}(\gamma(s))
{\rm Tr}[U(0,s)]\tau^{k}U(s,1)],
\]
where we have written $\ut{M}(\gamma(s))\sqrt{\gamma(s)}$ as a scalar
function
$M(\gamma(s))$. Here we see that simply by multiplying with $\epsilon$ and
performing the limit $ \epsilon\rightarrow0$ we get a finite, well defined
result:

\[
\lim_{\epsilon\rightarrow 0}{\epsilon\hat
{\cal H}_{\epsilon}(\ut{M})W_{\gamma}^{\rm  {smooth}}(A)}=
\]

\begin{equation}\label{smooth}
{3 Z\over 2\pi}\oint{ds}
\dot{\gamma}^{a}(s)\ddot{\gamma}^{b}(s)M(\gamma(s)) F_{ab}^{k}(\gamma(s)){\rm
Tr}[U(0,s)\tau^{k}U(s,1)] ,
\end{equation}
where $Z$ is an arbitrary renormalization constant. This result can be easily
generalized for the case of a loop $\gamma$ with kinks and/or self-
intersections - the closed integral in (\ref{smooth}) should be replaced by a
sum of integrals along smooth portions of the loop.

The result we obtained is different in its detail from the results obtained
previously (\cite{Smo1}, \cite{Blen}) because of the different regularization
schemes used, but in its general features it is similar. Unfortunately our
result faces the same problem encountered before - it is background dependent
because of the presence of the ``acceleration" term $\ddot{\gamma}^{b}(s)$
(see \cite {Pull} and \cite {Smo3} for discussion).

\subsection{Contribution from self-intersections}
The second contribution in the action of the Hamiltonian constraint on an
Wilson loop will come from intersections. Let us consider a single self-
intersecting loop. Let $\vec {\eta}^{+}$ and $\vec {\eta}^{-}$ be the unit
tangent vectors to the loop at the intersection:

\begin{figure}
\par
\centerline{
\epsfig{figure=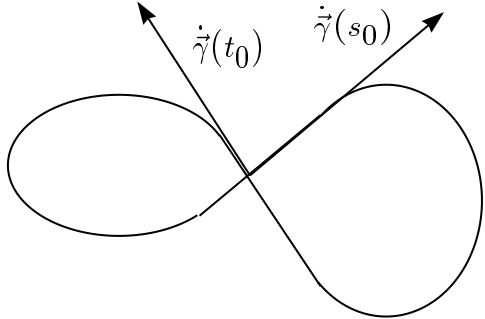,width=70mm}
}
\par
\caption{Loop with self-intersection.}
\label{f2}
\end{figure}

\begin{figure}
\par
\centerline{
\epsfig{figure=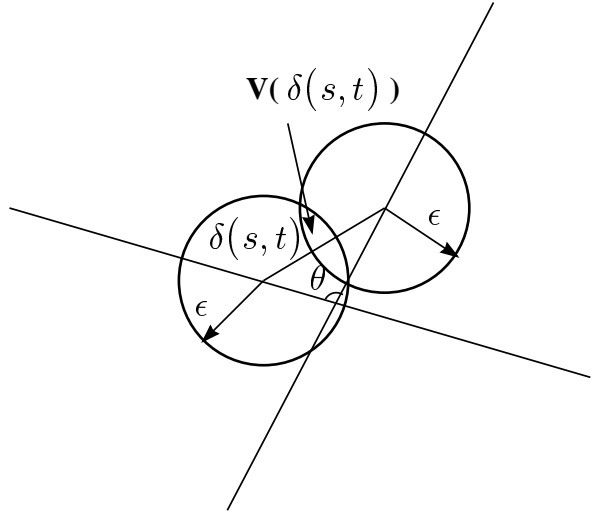,width=80mm}
}
\par
\caption{Integration for the loop with self-intersection.}
\label{f3}
\end{figure}

Here we will have two separate cases - one with $ \dot{\vec \gamma} (s_{0})
\equiv \vec{\eta}^{+}$ and $ \dot{\vec \gamma}(t_{0}) \equiv \vec{\eta}^{-}$,
and another one with $s$ and $t$ replaced. In the above relations $s_{0}$ and
$t_{0}$ are the values of the parameters along the loop, corresponding to the
intersecting point. Again we use a parametrization in which $|\dot{\vec \gamma}
(s_{0})|=\dot{\vec \gamma} (t_{0})=1$. Keeping in mind the limiting procedure
we are performing, we can write the result from the action of the Hamiltonian
constraint on the intersecting portion of the loop as:

\[
\hat{\cal H}_{\epsilon}(\ut{M})W_{\gamma}^{int}(A) =  {9 \over
16\pi^{2}\epsilon^{6}}
\ut{M}(\gamma(s_{0})) \epsilon_{ijk}
F_{ab}^{k}(\gamma(s_{0}))\sqrt{h(\gamma(s_{0}))} \]

\[
\oint{ds}\oint{dt}
\dot{\gamma}^{a}(s)\dot{\gamma}^{b}(t)L^{ij}(s,t;\gamma)
\int{d^{3}x}\sqrt{h(x)}
\vartheta [ \epsilon - |\vec{x} - \vec{\gamma(s)}|]\vartheta [ \epsilon -
|\vec{x} - \vec{\gamma(t)}|].
\]
This last expression can be further simplified by replacing the arguments of
the holonomies in $L^{ij}(s,t;\gamma)$ by the values of the parameters $s$ and
$t$ corresponding to the intersection.

\[
\hat{\cal H}_{\epsilon}(\ut{M})W_{\gamma}^{int}(A) =  {9 \over
16\pi^{2}\epsilon^{6}}
{\epsilon}_{ijk}\ut{M}(\gamma(s_{0})) F_{ab}^{k}(\gamma(s_{0}))
\sqrt{h(\gamma(s_{0}))}
\]

\[
\{ ({\eta}^{+})^{a}({\eta}^{-})^{b} {\rm Tr}[U(0,t_{0}){\tau}^{j}U(t_{0},s_{0})
{\tau}^{i}U(s_{0},1)] +
\]

\[
+ ({\eta}^{+})^{b}({\eta}^{-})^{a} {\rm Tr}[U(0,s_{0}){\tau}^{i}U(s_{0},t_{0})
{\tau}^{j}U(t_{0},1)]\}
\]

\begin{equation}\label{Hpsi4}
\oint{ds} \oint{dt}
\int{d^{3}x}\sqrt{h(x)}
\vartheta [ \epsilon - |\vec{x} - \vec{\gamma(s)}|]\vartheta [ \epsilon -
|\vec{x} - \vec{\gamma(t)}|].
\end{equation}
To perform the integration we can again make use of (\ref {Volume}) and write
the last 5-fold integral as:

\[
I(\epsilon,\theta)=\int{ds}\int{dt}V(\delta(s,t)),
\]
where the limits of integration are to be determined from Figure \ref{f3}.

After tedious but straightforward calculations (see Appendix 2) we get the
following result:

\[
\hat{\cal H}_{\epsilon}(\ut{M})W_{\gamma}^{int}(A)={69 \over 20\epsilon
\sin{\theta}}{\epsilon}_{ijk}\ut{M}(\gamma(s_{0})) F_{ab}^{k}(\gamma(s_{0}))
\sqrt{h(\gamma(s_{0}))}
\]

\[
 ({\eta}^{+})^{a}({\eta}^{-})^{b} {\rm Tr}[U(t_{0},s_{0}){\tau}^{i}U(s_{0},
t_{0}){\tau}^{j}] + {\cal O}(1),
\]
where $\theta$ is the angle between the vectors $\vec {\eta}^{+}$ and
$\vec {\eta}^{-}$.
Again after performing multiplicative renormalization  we get:

\begin{equation}\label{Renint}
\lim_{\epsilon\rightarrow 0}\epsilon\hat{\cal H}_{\epsilon}(\ut{M})
W_{\gamma}^{int}(A)=
\]
\[
={69{\epsilon}_{ijk}\over 20\sin{\theta}}M^{int} (F_{ab}^{k})^{int}
 ({\eta}^{+})^{a}({\eta}^{-})^{b} {\rm Tr}[U(t_{0},s_{0}){\tau}^{i}U(s_{0},
t_{0}){\tau}^{j}],
\end{equation}
where $M^{int}$ and $(F_{ab}^{k})^{int}$ are the values of $M(x)$ and
$F_{ab}^{k}(x)$ at the intersection point.
To make this result easier to understand let us write:

\[
F_{ab}^{k}={1 \over 2}{\epsilon}_{abc}B^{ck},
\]
where $B^{ck}$ is the ``magnetic field". Then we will have:

\[
({\eta}^{+})^{a}({\eta}^{-})^{b}{\epsilon}_{abc}B^{ck}=
(\vec{\eta}^{+}\times \vec{\eta}^{-})_{c}B^{ck}=
\hat{n}_{c}B^{ck}\sin{\theta},
\]
where $ {\hat n}_{c}$ is the unit vector, normal to the plane defined by the
loop at the intersection.Thus finally we get:

\begin{equation}\label{finint}
\lim_{\epsilon\rightarrow 0}\epsilon\hat{\cal
H}_{\epsilon}(\ut{M})W_{\gamma}^{int}
(A)={69{\epsilon}_{ijk}Z\over 40}M^{int} (\hat{n}_{c}B^{ck})^{int}
{\rm Tr}[U(t_{0},s_{0}){\tau}^{i}U(s_{0},t_{0}){\tau}^{j}].
\end{equation}
In this form it is clear that the result is independent from the metric used
in the regularization. Thus (up to a numerical factor) we recover the same
result like the one obtained in \cite{SmoRov} in the loop representation.

\subsection{Contributions from kinks}
The case  of a loop with a kink is similar to the one with an intersection.
The final result after performing the integration and renormalization is:

\begin{equation}\label{kink}
\lim_{\epsilon\rightarrow 0}\epsilon\hat{\cal H}_{\epsilon}(\ut{M})
W_{\gamma}^{kink}(A)={f(\theta) \over \sin{\theta}}M^{kink} (F_{ab}^{k})^{kink}
 ({\eta}^{+})^{a}({\eta}^{-})^{b} {\rm Tr}[U(0,s_{0}){\tau}^{k}U(s_{0},1)]
\end{equation}
where

\[
f(\theta)={3\over 40}\left( 23\left( 1-{\theta \over \pi} \right) - {9 \over
\pi}\sin{\theta}\cos{\theta }\right)
\]
Thus when there is a kink on the loop the result is again background dependent.
The dependance shows up in the presence of $f(\theta)$ - a function of the
angle between the tangent vectors at the kink, for the definition of which we
need a background metric.

\section{The algebra of the constraints}

In this section we will present the calculations of the constraint algebra
with a Regularized Hamiltonian constraint. Tsamis and Woodard suggest in
\cite{Tsamis} that a regulating procedure, which is not coordinate invariant
should destroy the algebra closure. Surprisingly this is not exactly the case.
After appropriate redefinition of the regulators the Hamiltonian and
Diffeomorphism constraints do close, which means that the evolution generated
by the regulated Hamiltonian constraint is
consistent with the requirement for coordinate invariance.

\subsection{Hamiltonian with Gauss constraint}
We start with the Poisson bracket between the Regularized Hamiltonian and the
Gauge constraints:

\[
\{{\cal G}({\bf \vec{N}}), {\cal H}_{\epsilon}(\ut{M})\} =
\]

\[
{1 \over 2}\int d^{3}x
\epsilon_{ijk}\ut{M}(x)\int d^{3}y  f_{\epsilon}(\tilde{x},y)\int d^{3}z
f_{\epsilon}(\tilde{x},z)
\{ {\cal G}({\bf \vec{N}}),F_{ab}^{k}(x)
\tilde{E}^{ai}(y)\tilde{E}^{bj}(z)\}=
\]

\[
\int d^{3}x
\ut{M}(x)\int d^{3}y  f_{\epsilon}(\tilde{x},y)\int d^{3}z  f_{\epsilon}
(\tilde{x},z)F_{ab}^{k}(x)\tilde{E}^{bj}(z)
\]

\begin{equation}\label{GHr}
[\left( {N}_{j}(x) - {N}_{j}(y)\right){E}^{a}_{k}(y) + \left( {N}_{k}(y) -
{N}_{k}(z)\right){E}^{a}_{j}(y)].
\end{equation}

In the unregulated case this bracket gives zero. Before considering the
obtained result as a problem let us remember that the above calculation is a
part of a limiting procedure. Because of the terms $\left( {N}_{j}(x) -
{N}_{j}(y)\right)$ and $\left( {N}_{k}(y) - {N}_{k}(z)\right)$ if we perform
the limit $\epsilon\rightarrow0$ at a classical level, the expression
(\ref{GHr}) will be of order $\epsilon$ and will vanish. In the quantum case,
which is interesting for us, the result is similar. Both terms in (\ref{GHr})
have space indices like in a Hamiltonian constraint but rearranged internal
indices. This means that the final result of the action of $\{{\cal G}({\bf
\vec{N}}), {\cal H}_{\epsilon}(\ut{M})\}$ on Wilson loops will contain
different
types of braking of the loop $\gamma$ but it will be ${\cal O}(1)$. Thus we
will have:

\[
\lim_{\epsilon\rightarrow 0}\epsilon\{{\cal G}({\bf \vec{N}}), {\cal H}_
{\epsilon}(\ut{M})\}W_{\gamma}(A)=0.
\]

\subsection{Hamiltonian with Diffeomorphism constraint}
To calculate the Poisson bracket between the Hamiltonian and the Diffeomorphism
constraints we will use (\ref {Lf}) to get:

\[
\{{\cal H}_{\epsilon}^{ff}(\ut{M}),{\cal C}(\vec{N})\} =
\]

\[= {1 \over 2}\int d^{3}x
\epsilon_{ijk}\ut{M}(x)\int d^{3}y  f_{\epsilon}(\tilde{x},y)\int d^{3}z
f_{\epsilon}(\tilde{x},z)
\{ F_{ab}^{k}(x)\tilde{E}^{ai}(y)\tilde{E}^{bj}(z),
{\cal C}(\vec{N})\} =
\]

\[
= {1 \over 2}\int d^{3}x\epsilon_{ijk}\ut{M}(x)\int d^{3}y
f_{\epsilon}(\tilde{x},
y)\int d^{3}zf_{\epsilon}(\tilde{x},z) [{\cal L}_{\vec{N}}
F_{ab}^{k}(x)\tilde{E}^{ai}(y)\tilde{E}^{bj}(z) +
\]

\begin{equation}\label{HffC}
+ F_{ab}^{k}(x){\cal L}_{\vec{N}} \tilde{E}^{ai}(y)\tilde{E}^{bj}(z)
+ F_{ab}^{k}(x)\tilde{E}^{ai}(y)
{\cal L}_{\vec{N}}\tilde{E}^{bj}(z) ].
\end{equation}
In this case it is important for us to use more explicit notation - in
${\cal H}_{\epsilon}^{ff}$ we have shown explicitly the type of regulating
functions we have used.
As usual we are working on a compact manifold which will allow us to do
integration by parts disregarding all boundary terms. Performing such an
integration in (\ref{HffC}), we get:

\[
\{{\cal H}_{\epsilon}^{ff}(\ut{M}),{\cal C}(\vec{N})\} =
\]

\[
= - {1 \over 2}\int d^{3}x \int d^{3}y \int d^{3}z \epsilon_{ijk} F_{ab}^{k}(x)
\tilde{E}^{ai}(y) \tilde{E}^{bj}(z)
\{{\cal L}_{(x)\vec{N}}\left( \ut{M}(x)  f_{\epsilon}(\tilde{x},y)f_{\epsilon}
(\tilde{x},z)  \right) +
\]

\[
+ \ut{M}(x)\left({\cal L}_{(y)\vec{N}}f_{\epsilon}(\tilde{x},y) \right)
f_{\epsilon}(\tilde{x},z)+ \ut{M}(x)  f_{\epsilon}(\tilde{x},y)
\left({\cal L}_{(z)\vec{N}} f_{\epsilon}(\tilde{x},z)  \right) \}.
\]
After rearranging the terms in the last expression we will get:

\[
\{{\cal H}_{\epsilon}^{ff}(\ut{M}),{\cal C}(\vec{N})\} =
\]

\[
= - {\cal H}_{\epsilon}^{ff}({\cal L}_{\vec{N}}\ut{M}) - {1 \over 2}\int
d^{3}x
\int d^{3}y \int d^{3}z \epsilon_{ijk} \ut{M}(x){\partial}_{c}{\tilde
{N}}^{c}(x) F_{ab}^{k}(x) \tilde{E}^{ai}(y) \tilde{E}^{bj}(z)
\]

\[
 f_{\epsilon}(\tilde{x},z) 2\left\{  f_{\epsilon}(\tilde{x},y) + {h(x) \over
{ \partial_{a}}N^{a}(x)}\left( {N}^{c}(x){\partial \over \partial {x}^{c}}
f_{\epsilon}(x,y) + {N}^{c}(y){\partial \over \partial {y}^{c}} {f_{\epsilon}
(x,y)} \right)\right\}.
\]
This final expression can be written in the form:

\begin{equation}\label{result}
\{{\cal H}_{\epsilon}^{ff}(\ut{M}),{\cal C}(\vec{N})\} = - {\cal
H}_{\epsilon}^{ff}
({\cal L}_{\vec{N}}\ut{M}) + {\cal
H}_{\epsilon}^{fg}(2\ut{M}{\partial}_{c}{\tilde
{N}}^ {c}) - {\cal H}_{\epsilon}^{ff}(2\ut{M}{\partial}_{c}{\tilde {N}}^{c})
\end{equation}
In the second term in the right hand side of (\ref{result}) the superscript
$g$ stands for the  expression:

\[
g_{\epsilon}(\tilde{x},y) = - {h(x) \over { \partial_{a}}N^{a}(x)}
\left( {N}^{c}(x){\partial \over \partial {x}^{c}} f_{\epsilon}(x,y) +
{N}^{c}(y){\partial \over \partial {y}^{c}} {f_{\epsilon}(x,y)} \right).
\]
It can be shown that the function $g_{\epsilon}(\tilde{x},y) $ satisfies the
requirement

\begin{equation}
\lim_{\epsilon\rightarrow 0}{\int{d^{3}x \phi (x)g_{\epsilon}(\tilde{x},y)}}
= \phi (y).
\end{equation}
for any smooth function $\phi(x)$ so it can be used as a regulating function.
This means that the Poisson bracket between Hamiltonian and Diffeomorphism
constraints gives as a result a sum of three regulated Hamiltonian constraints.
Thus the process of regularization changes the unregulated expression
(\ref{C,H}) in such a way so this change does not destroy the constraints
closure. Obviously if we perform in (\ref{result}) the limit
$\epsilon\rightarrow 0$ the last two terms we cancel each other and thus we
will recover the unregulated result (\ref{C,H}). If we proceed and quantize,
the expression (\ref{result}) will transform into its quantum version but the
closure will not be affected in the process of quantization and it still will
hold.

\subsection{Two regulated Hamiltonian constraints}

When calculating the Poisson bracket of two regulated Hamiltonian constraints
we have to keep in mind the fact that the quantum commutator of two
unregulated Hamiltonian constraints does not give combination of constraints.
This means that it makes sense for us to work only at the classical level of
the theory. Classically the Poisson brackets of two regulated Hamiltonian
constraints can be shown to give:

\[
\{{\cal H}_{\epsilon}(\ut{M}),{\cal H}_{\epsilon}(\ut{N})\}=
\]

\[
\int{d^{3}x}\int{d^{3}y}\int{d^{3}z}\int{d^{3}x'}
\int{d^{3}z'}f_{\epsilon}(\tilde{x'},x)f_{\epsilon}(\tilde{x'},z')
f_{\epsilon}(\tilde{x},y)f_{\epsilon}(\tilde{x},z)
\]

\[
[\ut{M}(x')\partial_{a}\ut{N}(x)-\ut{N}(x')\partial_{a}\ut{M}(x)]\tilde{E}^{a}_{n}(y)
\tilde{E}^{dn}(z')\tilde{E}^{b}_{k}(z)F_{bd}^{k}(x') +
\]

\[
+ \int{d^{3}x}\int{d^{3}y}\int{d^{3}z}\int{d^{3}x'}
\int{d^{3}z'}f_{\epsilon}(\tilde{x'},x)f_{\epsilon}(\tilde{x'},z')
F_{bd}^{k}(x')\tilde{E}^{dn}(z')
\]

\[
\{[\ut{N}(x')\partial_{a}\ut{M}(x)-\ut{M}(x')\partial_{a}\ut{N}(x)]
f_{\epsilon}(\tilde{x},y)f_{\epsilon}(\tilde{x},z)
\tilde{E}^{a}_{k}(y)
\tilde{E}^{b}_{n}(z) +
\]

\[
+ [\ut{N}(x')\ut{M}(x) - \ut{M}(x')\ut{N}(x)]\tilde{E}^{a}_{[k}(y)
\tilde{E}^{b}_{n]}(z)
{\partial \over \partial x^{a}}{(f_{\epsilon}(\tilde{x},y)
f_{\epsilon}(\tilde{x},z))} +
\]

\begin{equation}\label{HregH}
+ i\epsilon_{kmn}[\ut{N}(x')\ut{M}(x) -
\ut{M}(x')\ut{N}(x)]f_{\epsilon}(\tilde{x},y) f_{\epsilon}(\tilde{x},z)
A_{ap}(x)\tilde{E}^{a[p}(y)
\tilde{E}^{bm]}(z)\}.
\end{equation}

In the last messy expression the first 5-fold integral can be rearranged in
the form: \\

\begin{equation}\label{regK}
\int{d^{3}x}\int{d^{3}x'}K_{\epsilon}^{d}(\tilde{x'},x)\int{d^{3}z}
f_{\epsilon}(\tilde{x},z)\tilde{E}^{b}_{k}(z)F_{bd}^{k}(x')
\end{equation}
where $K_{\epsilon}^{d}(\tilde{x'},x)$ is the smeared version of (\ref {K}):

\[
K_{\epsilon}^{d}(\tilde{x'},x)=f_{\epsilon}(\tilde{x'},x)
[\ut{M}(x')\partial_{a}\ut{N}(x)-\ut{N}(x')\partial_{a}\ut{M}(x)]
\]

\[
\int{d^{3}y}f_{\epsilon}(\tilde{x},y)
\tilde{E}^{a}_{n}(y)
\int{d^{3}z'}f_{\epsilon}(\tilde{x'},z')\tilde{E}^{dn}(z').
\]

It is straightforward to check that classically for any smooth function
$\phi(x)$ $K_{\epsilon}^{d}(\tilde{x'},x)$ behaves like are regulating
function:
\[
\lim_{\epsilon\rightarrow 0}{\int{{d}^{3}x'K_{\epsilon}^{d}(\tilde{x'},x)
\phi(x')}}=K^{d}(x)\phi(x).
\]
If we classically perform the limit $\epsilon\rightarrow 0$ in (\ref{HregH})
as
we should expect, it reduces to the unregulated expression (\ref{H,H}). The
terms in the second 5-fold integral in (\ref{HregH}) vanish - the first one
because of the antisymmetry of $F_{bd}^{k}(x)$, the second and the third one
- because of the expression
$[\ut{N}(x')\partial_{a}\ut{M}(x)-\ut{M}(x')\partial_{a}\ut{N}(x)]$.
But with the presence of $\epsilon$ we are not able to write (\ref{HregH}) as a
combination of constraints. Thus in the process of regularization the
constraint closure is lost because of the $\{{\cal H}_{\epsilon}(\ut{M}) ,
{\cal H}_{\epsilon}(\ut{N})\}$ bracket, but this most probably just reflects
the
fact that the corresponding unregulated expression also prevents the quantum
algebra from closing.

\section{Conclusions}
In conclusion we would like to emphasize again on the main problems we have
encountered in our work:
\begin{itemize}
\item
The first set of problems arises in connection with the requirement for
background independence of the action of the regulated Hamiltonian constraint.
The result we obtained shows that the arbitrary metric we have used in the
calculations survives when the Hamiltonian acts on smooth portions of loops and
on loops with kinks. In the case of loops with self-intersections the
background dependance drops completely. In the case of smooth portions of loops
the troublesome term is the so called ``acceleration term". As pointed out in
\cite
{Pull} the ``acceleration term" $\ddot{\gamma}^{b}(s)$ is not a tensor
quantity and its presence in (\ref {smooth}) means that the result depends on
the arbitrary metric we have used. This problem has been discussed in
\cite{Rov1} where it is shown that smearing of the loops on which the loop
operators in the loop representation are based makes the ``acceleration
term" to vanish. However the
corresponding solution in the connection representation requires further
investigation.

On the other hand the problem with the loops with kinks probably requires
appropriate redefinition of the regularization procedure so the background
dependant factor gets absorbed in the process of regularization \cite {Smo4} .
To summarize: Even if the smearing of the loop
succeeds to remove the background dependance coming from the smooth portions
of the loop, we have to face the problem of the imprint of the metric used in
the case of a loop with a kink.
\item
Another, though similar problem is concerned with the constraints closure. The
Poisson bracket between two regulated Hamiltonian constraints give as a result
an expression which apparently can not be cast into the form of a sum of
(regulated ) constraints. It could be the case that the quantum commutator will
preserve the constraints closure but there is no obvious reason for this to
happen. It is possible that the problem should be faced at an earlier stage - a
factor ordering should be sought, which preserves both the constraints closure
and the physical meaning of the constraints.
\end{itemize}

\section{Acknowledgments}
I am grateful to Jorge Pullin for giving me the idea for this work and for the
discussions afterwards. I also thank Plamen Fiziev, Don Neville, Petko Nikolov,
Lee Smolin, and Modhavan Varadarajan for comments, suggestions and criticisms.
\newpage
\section{Appendices}
\subsection{Appendix 1}
Here we will show the basic steps in the calculation of (\ref{Hpsi3}). We
start from:
\[
\hat{\cal H}_{\epsilon}(\ut{M})W_{ \gamma}^{\rm smooth}(A) =  {9 \over
16\pi^{2} \epsilon^{6}}\oint{ds} \oint{dt}
\dot{\gamma}^{a}(s)\ddot{\gamma}^{b}(s)(t-s)L^{ij}(s,t;\gamma)
\]

\[
\ut{M}(x_{0})F_{ab}^{k}(x_{0})\sqrt{h(x_{0})} V(\delta(s,t)).
\]
First, from the expansion of $L^{ij}(s,t;\gamma)$ we will get:
\[
L^{ij}(s,t;\gamma)={\rm Tr}[ \vartheta(s-t)U(0,s)\tau^{j}\tau^{i}U(s,1) +
\vartheta(t-s)U(0,s)\tau^{i}\tau^{j}U(s,1)]=
\]
\[
={1 \over 2} {\epsilon}^{jik}{\rm Tr}[ \vartheta(s-t)U(0,s)\tau_{k}U(s,1) -
\vartheta(t-s)U(0,s)\tau_{k}U(s,1)].
\]
In the function $V(\delta(s,t))$ working to lowest order in $\epsilon$ we can
replace $\delta$ with $|t-s|$. Thus the integral reduces to:
\[
\hat{\cal H}_{\epsilon}(\ut{M})W_{ \gamma}^{\rm smooth}(A)=
\]
\[={3 \over 16\pi \epsilon^{3}}\oint{ds}
\dot{\gamma}^{a}(s)\ddot{\gamma}^{b}(s){\rm Tr}[ U(0,s)\tau_{k}U(s,1)]
\ut{M}({\gamma}(s))F_{ab}^{k}({\gamma}(s))\sqrt{{\gamma}(s))}
\]
\[
\{ \int_{0}^{2\epsilon{\delta}^{+}}{(t-s)\left[2 - {3(t-s) \over 2\epsilon} +
\left({t-s \over 2\epsilon}\right)^{3}\right]}d(t-s) -
\]
\[
 \int_{-2\epsilon{\delta}^{-}}^{0}{(t-s)\left[2 + {3(t-s) \over 2\epsilon} -
\left({t-s \over 2\epsilon}\right)^{3}\right]}d(t-s)\}=
\]
\[={3 \over 4\pi \epsilon}\oint{ds}
\dot{\gamma}^{a}(s)\ddot{\gamma}^{b}(s){\rm Tr}[ U(0,s)\tau_{k}U(s,1)]
\ut{M}({\gamma}(s))F_{ab}^{k}({\gamma}(s))\sqrt{{\gamma}(s))}
\]
\[
\left\{  ({\delta}^{+})^{2} - ({\delta}^{+})^{3} + {1 \over 5}
{({\delta}^{+})}^{5} + ({\delta}^{-})^{2} + ({\delta}^{-})^{3} -
{1 \over 5}{({\delta}^{-})}^{5}
 \right\}.
\]
Because $(({\delta}^{+})^{3} - ({\delta}^{-})^{3} )={\cal O}(\epsilon)$ and
also $(({\delta}^{+})^{5} - ({\delta}^{-})^{5} )={\cal O}(\epsilon)$ we can
write the action of the regulated Hamiltonian constraint on smooth portions
in the lowest order of $\epsilon$ as:
\[
\hat{\cal H}_{\epsilon}(\ut{M})W_{ \gamma}^{\rm smooth}(A)=
\]
\[
{3 \over 2\pi\epsilon}((\delta^{+})^{2}+(\delta^{-})^{2})\oint{ds}
\dot{\gamma}^{a}(s)\ddot{\gamma}^{b}(s)M(\gamma(s)) F_{ab}^{k}(\gamma(s)){\rm
Tr}[U(0,s)]\tau^{k}U(s,1)].
\]

\subsection{Appendix 2}
To obtain the contribution from self-intersections, we have to compute the
integral:

\[
I(\epsilon,\theta)=\int{ds}\int{dt}V(\delta(s,t))
\]
where $V(\delta(s,t))$ is given by \ref{Volume} and the limits of integration
can be determine from Figure 2. First we will change the variable $s$ into:
\[
\xi={s\sin{\theta} \over 2\epsilon}
\]
where $\sin{\theta}$ is the angle between $\vec{{\eta}}^{+}$ and
$\vec{{\eta}}^{-}$. Also we will introduce another variable $\varphi$ via the
relation $t=2\epsilon\xi\tan{\varphi}$. Thus we will have:

\[
V(\xi,\varphi)={2\pi{\epsilon}^{3} \over 3}\left[ 2 - {3\xi \over
\cos{\varphi}} +{ {\xi}^{3} \over {\cos^{3}{\varphi}}}\right]
\]

Keeping $\xi$ fixed we can perform the integration with respect to $\varphi$
to get:
\[
{8\pi{\epsilon}^{4} \xi\over 3}\int_{0}^{\arccos{\xi}}{{d\varphi \over
{\cos}^{2}{\varphi}}\left[2 - {3\xi \over \cos{\varphi}} +{ {\xi}^{3} \over
{\cos^{3}{\varphi}}}\right]} =
\]

\[
=\pi{\epsilon}^{4}\left\{ (4 + {\xi}^{2})\sqrt{1- {\xi}^{2}} - {1 \over 2}
{\xi}^{2}(4 - {\xi}^{2})\ln{\left[{1 + \sqrt{1- {\xi}^{2}}  \over 1 -
\sqrt{1- {\xi}^{2}}}\right]}\right\}.
\]

To obtain the final result we have to perform the integration with respect to
$\xi$:

\[
I(\epsilon,\theta)={4\pi{\epsilon}^{5} \over \sin{\theta}}\int_{0}^{1}{d\xi}
\left\{ (4 + {\xi}^{2})\sqrt{1- {\xi}^{2}} - {1 \over 2}{\xi}^{2}(4 -
{\xi}^{2})
\ln{\left[{1 + \sqrt{1- {\xi}^{2}}  \over 1 - \sqrt{1- {\xi}^{2}}}\right]}
\right\}.
\]

Thus finally after this exercise in calculus we get:

\[
I(\epsilon,\theta)={4\pi^{2}{\epsilon}^{5} \over \sin{\theta}}{ 23 \over 30}.
\]

\end{document}